\documentclass{elsart}

 \usepackage{epsfig}

\usepackage{amssymb}
\usepackage{multirow}
\usepackage{textcomp}
\usepackage{ulem}
\begin{document}

\begin{frontmatter}

% Title, authors and addresses

% use the thanksref command within \title, \author or \address for footnotes;
% use the corauthref command within \author for corresponding author footnotes;
% use the ead command for the email address,
% and the form \ead[url] for the home page:
% \title{Title\thanksref{label1}}
% \thanks[label1]{}
% \author{Name\corauthref{cor1}\thanksref{label2}}
% \ead{email address}
% \ead[url]{home page}
% \thanks[label2]{}
% \corauth[cor1]{}
% \address{Address\thanksref{label3}}
% \thanks[label3]{}

%\title{Study of the $\beta$-delayed $^{11}$Li charged particle emission }
\title{Study of $\beta$-delayed 3-body and 5-body breakup channels observed in the decay of $^{11}$Li}
% use optional labels to link authors explicitly to addresses:
% \author[label1,label2]{}
% \address[label1]{}
% \address[label2]{}

\author[csic]{M. Madurga},
\author[csic]{M.J.G. Borge\corauthref{cor}},
\author[lpc]{J.C. Angelique},
\author[goteborg]{L. Bao},
\author[isolde]{U. Bergmann},
\author[lpc,romania]{A. But\u{a}},
\author[isolde]{J. Cederk\"{a}ll},
\author[aarhus]{C.Aa. Diget},
\author[isolde]{L.M. Fraile\thanksref{nowfraile}},
\author[aarhus]{H.O.U. Fynbo},
\author[aarhus]{H.B. Jeppesen\thanksref{nowhenrik}},
\author[goteborg]{B. Jonson},
\author[lpc]{F. Mar\'echal\thanksref{nowf}},
\author[lpc]{F.M. Marqu\'es},
\author[goteborg]{T. Nilsson},
\author[goteborg]{G. Nyman},
\author[lpc]{F. Perrot\thanksref{nowf}},
\author[aarhus]{K. Riisager},
\author[csic]{O. Tengblad},
\author[goteborg]{E. Tengborn},
\author[csic]{M. Turri\'on},
\author[goteborg]{K. Wilhelmsen}

\corauth[cor]{Corresponding author.}

\thanks[nowfraile]{\scriptsize {Present address:} \tiny{Departamento de F\'isica At\'omica, Molecular y Nuclear,  Universidad  Complutense, E-28040 Madrid, Spain} }
\thanks[nowhenrik]{\scriptsize {Present address:} \tiny{Nuclear Science Division, Lawrence Berkeley National Laboratory, CA-94720
Berkley, USA} }
\thanks[nowf]{\scriptsize{Present address:} \tiny{Institut de Recherches Subatomiques, IN2P3-CNRS, F-67037 Strasbourg Cedex 2, France} }
\address[csic]{Instituto de Estructura de la Materia, CSIC, E-28006 Madrid, Spain}
\address[lpc]{Laboratoire de Physique Corpusculaire, IN2P3-CNRS, ISMRA, Universit\'{e} de Caen, Boulevard Mar\'{e}chal Juin, 14050 CAEN Cedex, France}
\address[goteborg]{Fundamental Physics, Chalmers University of Technology, S-41296 G\"oteborg, Sweden}
\address[isolde]{PH Department, CERN, CH-1211 Gen\`eve, Switzerland}
\address[romania]{National Institute for Physics and Nuclear Engineering, P.O. Box MG-6, Bucharest, Romania}
\address[aarhus]{Department of Physics and Astronomy, University of Aarhus, DK-8000 {\AA}rhus, Denmark}

\begin{abstract}

{
The $\beta$-delayed charged particle emission from $^{11}$Li has been studied with emphasis on the three-body n$\alpha^6$He and five-body 2$\alpha$3n channels from the 10.59 and 18.15 MeV states in $^{11}$Be. Monte Carlo simulations using an R-matrix formalism lead to the conclusion that the $^A$He resonance states play a significant role in the break-up of these states. The results exclude an earlier assumption of a phase-space description of the break-up process  of the 18.15 MeV state. Evidence for extra sequential decay paths is found for both states. }
\end{abstract}

\end{frontmatter}

% main te. xt
\section{Introduction}
\label{intro}

The nucleus $^{11}$Li has provided many challenges to nuclear physicists over the last decades, see e.g. \cite{Jon04} for a recent overview. Its beta-decay presents one of these challenges that have not yet been fully solved. The large Q-value (20.68 MeV) for the $\beta$ decay and the low particle separation energies in the daughter $^{11}$Be implies that many decay channels are open. The most extreme channel has a five-body final state consisting of two alpha particles and three neutrons. Decays into this channel, a miniature multifragmentation, and the fact that the constituents (n and $\alpha$) are barely kept together allows structure interpretations in terms of cluster or few-body models to be tested. On the experimental side, a thorough investigation of the decay is only possible through the recording of the decay products in coincidence. This has so far only been reported in one earlier experiment \cite{langevin} that had a quite limited angular resolution.

The $^{11}$Li $\beta$-decay involves the largest number of decay  channels ever detected and experimental results have been reported for $\beta$n \cite{roeckl}, $\beta$2n \cite{azumab2n}, $\beta$3n \cite{azuma}, $\beta$t \cite{langevin2} and $\beta$d \cite{mukha}. The established channels involve the emission of $\alpha$ particles (2$\alpha$+3n), $^6$He ($^6$He+$\alpha$+n), tritons ($^8$Li+t), deuterons ($^9$Li+d) and the emission of 1n and 2n feeding states in $^{10}$Be and $^{9}$Be, respectively. The $\beta$-delayed charged particle emission of $^{11}$Li has been the subject of several previous studies \cite{langevin,langevin2,mukha,borge}.  The two channels involving coincident $\alpha$ particles were studied in one previous experiment \cite{langevin}. The 2$\alpha$3n channel was established to have contributions from the break-up of two states in $^{11}$Be, one at 10.59 MeV and the other at around 18.5 MeV. The $^6$He$\alpha$n channel was established as the $^6$He+$\alpha$ break-up of a state at about 9.4 MeV in $^{10}$Be fed in the neutron decay of the 10.59 MeV state in $^{11}$Be. 

By means of a gas-detector telescope for particle identification, the different $^6$He$\alpha$n, 2$\alpha$3n, $^{10}$Be+n, $^9$Be+2n, $^8$Li+t and $^9$Li+d decay channels of the $\sim$18.5 MeV state in $^{11}$Be and its branching ratios  were determined \cite{borge}. An R-matrix analysis established the energy of this state to be 18.15(15) MeV. The large reduced width of the $^9$Li+d channel with respect to the Wigner limit \cite{macfarlane} was interpreted as evidence of the $^{11}$Li(gs) $\beta$-decaying directly into the $^9$Li+d continuum.

A recent $\beta$-$\gamma$-n coincidence experiment using polarized $^{11}$Li  \cite{hirayama} determined the spin and parity of several levels fed in $^{11}$Be (E$<$11 MeV). In that work the  10.59 MeV level was found to contribute to the neutron spectrum via transitions to the $^{10}$Be(gs) and tentatively to the first 2$^+$ state at 3.4 MeV and to the unresolved doublet at $\sim$9.27 MeV  (4$^-$) and 9.4 MeV (2$^+$). The $\beta$-decay asymmetry of these three neutron transitions consistently determines the spin and parity of the 10.59 MeV level in $^{11}$Be as 5/2$^-$. Its 210(40) keV width was determined from the reaction $^9$Be(t,p)$^{11}$Be~\cite{selover}.

Additional information about the decay of the 10.59 and 18.15 MeV states in $^{11}$Be is obtained from three-neutron emission probability, which is equal to the  branching to the five body (2$\alpha$3n) final state. From the average branching ratio to the only bound state in $^{11}$Be, and the P$_{2n}$/P$_{1n}$, P$_{3n}$/P$_{1n}$ ratios determined by Azuma {\it et al.} \cite{azuma} a value of P$_{3n}$=1.9(2) \% is obtained.

The different experiments summarized above paint a fairly complicated picture of the $\beta$-delayed charged particle break-up of $^{11}$Li. It is neither well understood which levels in $^{11}$Be are contributing to the charged particle channels nor their break-up mechanisms. One should notice that significant $\beta$-decay strength in the 10-18 MeV energy range is predicted in shell model calculations \cite{Suz97}.

\begin{figure}
\centerline{\epsfig{figure=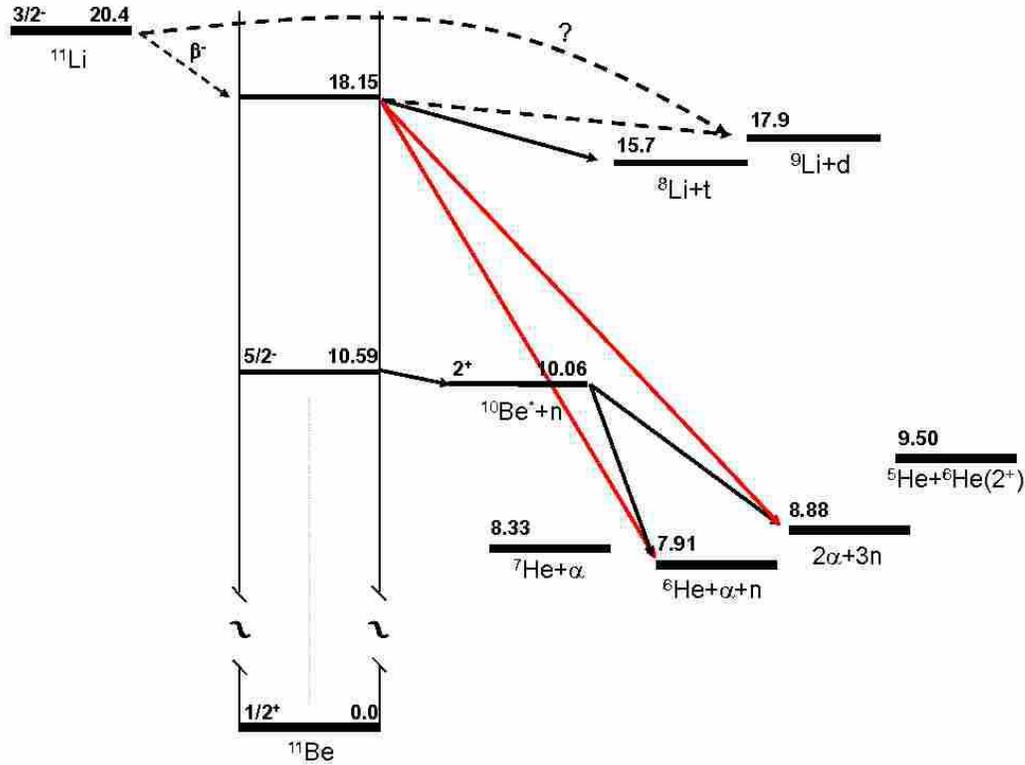,width=14cm}}
\caption{ Partial $^{11}$Li $\beta$-decay scheme of the $^{11}$Be levels above the charged particle threshold. The decay channels shown are taken from \cite{langevin,borge,tilley}. }
 \label{fig1}
\end{figure}

Fig. \ref{fig1} shows schematically the $^{11}$Li $\beta$-fed states in $^{11}$Be above the charged particle thresholds and the decay channels discussed above as established prior to this experiment. Since it is still not clear whether the $^9$Li+d channel proceeds sequentially through a state in $^{11}$Be at 18.15 MeV or directly from $^{11}$Li \cite{borge}, the figure displays both possibilities as dashed lines.

The main aim of the present paper is to provide significantly improved data on the $\beta$-delayed branches, especially those including two charged particles. These data allow a leap to be taken in the description of the break-up process following the weak decay. The interpretation presented here is based on known resonances in $^{11}$Be and intermediate nuclei in the decay chain. We consider not only decays through $^{10}$Be, but also channels involving $^A$He intermediate resonances. The following section describes the experiment, section 3 presents the analysis of the data, and the final section summarizes the results.

\section{Experiment}
\label{exp}

The experiment was carried out at the ISOLDE PSB Facility at CERN, where the $^{11}$Li activity was produced in fragmentation reactions in a Ta target irradiated with 1.4 GeV protons from the CERN PS Booster. The target container was connected to a surface ion source and the produced ions were accelerated to 30 keV. The  1$^+$ lithium  ions were mass selected from the different isobars  using the General Purpose Separator. The $^{11}$Li beam was subsequently brought to the center of the particle detector setup where it was stopped in a 60 $\mu$g/cm$^2$ carbon foil. The use of a thin foil, which minimizes the energy loss of the emitted charged particles, was possible due to the low acceleration voltage used in this experiment.

\begin{figure}
\centerline{\epsfig{figure=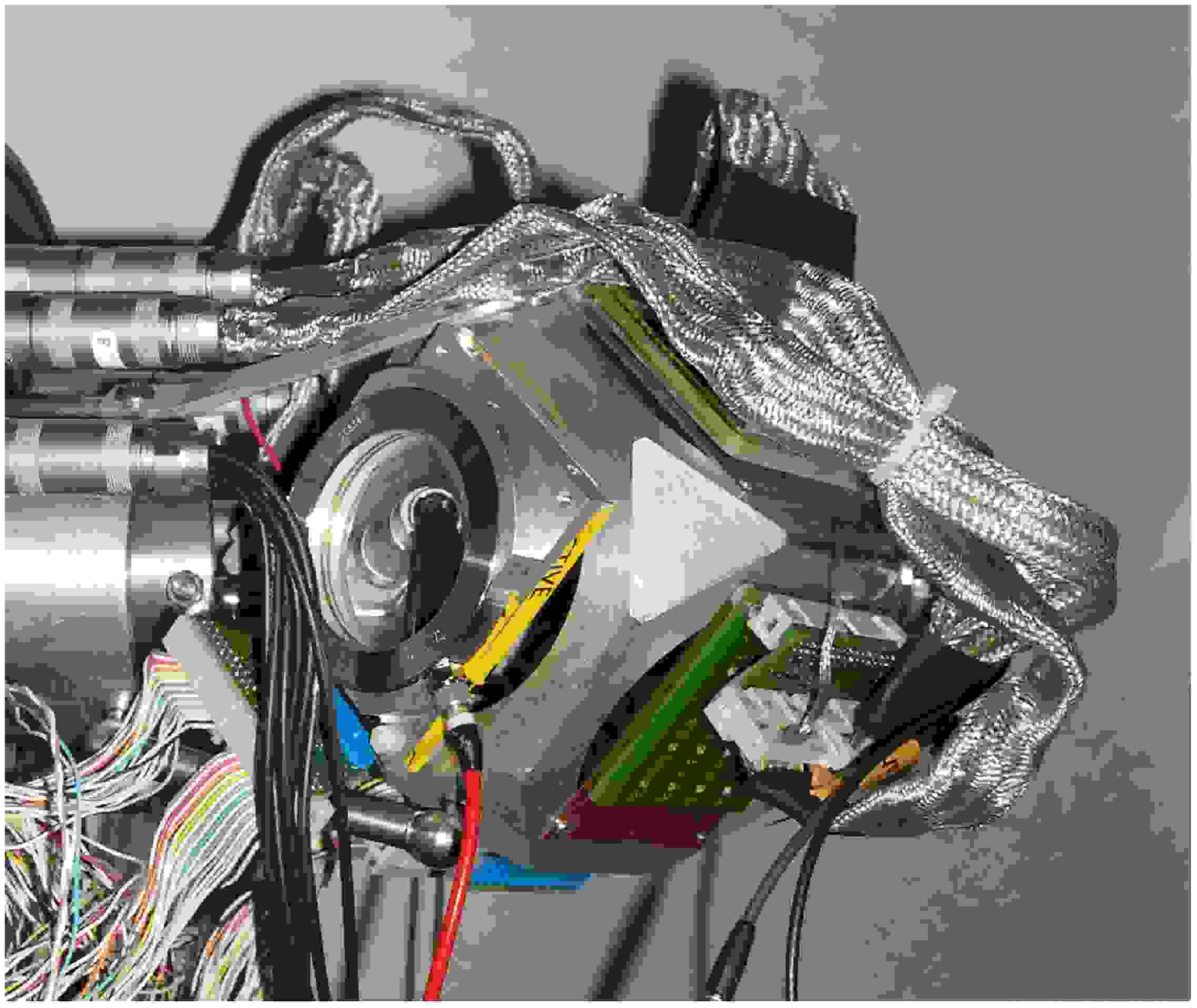,width=7cm}\epsfig{figure=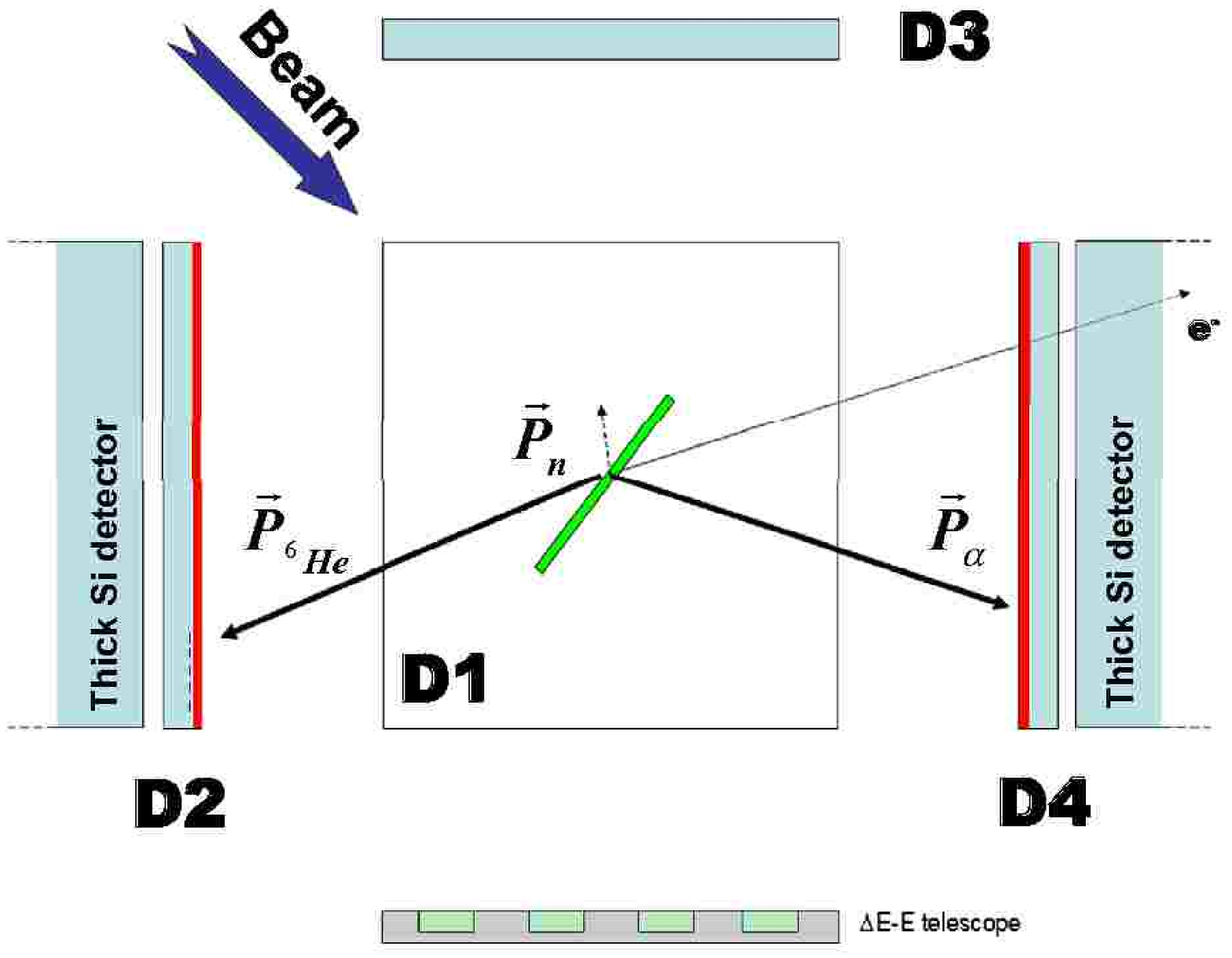,width=8cm}}
\caption{Left: A photo of the cubic frame where the DSSSD's were mounted. Right: schematic view of the detectors as they were placed in the cube.  A prototype telescope (not used in this analysis) is shown in the lower part of the drawing. The DSSSD D1 is on the picture plane. }
 \label{fig2}
\end{figure}

The setup, shown in Fig. \ref{fig2}, consisted of 4 particle telescopes with thin double-sided silicon strip  detectors (DSSSD) in front stacked with thick silicon pads, for $\beta$ detection. The telescopes were
mounted on the surfaces of a 10x10x10 cm$^3$ cubic frame, each covering 4\% of 4$\pi$.
Charge collection in these DSSSD's was carried out by 16 vertical and 16 horizontal strips,
constituting 256 detector pixels, with an angular resolution of $\sim$3.5$^o$. Each telescope combines large solid angle with high segmentation which gives large coverage with good angular resolution. The geometry of the setup defines three possible types of two-particle coincidences, depending on which detectors were hit. Hits in DSSSD D2 and DSSSD D4 (see Fig. \ref{fig2}), classified as \textit{180$^o$} coincidences, covered angles from  127$^o$~to 180$^o$~between the detected particles. Hits in DSSSD D3 and either DSSSD D2 or DSSSD D4, classified as \textit{90$^o$}~coincidences, covered angles from 37$^o$~to 143$^o$~between the detected particles. Finally, coincidences detected in the same detector, classified as \textit{0$^o$}~coincidences, covered angles from  0$^o$~to 50$^o$~between the detected particles. From here on we will concentrate on the \textit{180$^o$} coincidences, which have an efficiency estimated from the simulations of 5\% on average.

The silicon detectors were calibrated using standard $^{148}$Gd and triple alpha sources ($^{239}$Pu, $^{241}$Am and $^{244}$Cm) . Furthermore, the effective thickness of the carbon foil used as stopper was obtained from the analysis of the $\alpha$ particles coming from an online $^{20}$Na beam implanted in the foil before and after the experiment. The energy losses in non-active layers, such as the carbon-foil and the surface dead-layer of the silicon detectors are approximately 100 keV for $\alpha$ particles of energies around 1 MeV. Since 60\% of the $^{11}$Li $\beta$-delayed charged particles are expected to  be emitted with 1 MeV energy or less \cite{borge}, it is crucial to take these energy losses into account. The method developed within our collaboration \cite{uffed} takes advantage of the highly segmented nature of the DSSSD's to precisely define the path of the detected particle and thus allowing for the reconstruction of the energy losses in the non-active layers. The combined effect of the dead-layers and low energy noise gives detection thresholds around 180 keV.

\section{Analysis}

The use of highly segmented DSSSD detectors gives to this experiment a significant advantage compared to the previous studies of the $^{11}$Li $\beta$-delayed charged particle emission.  By using segmented detectors not only the energy of the particle is measured but also its direction with a good suppression of unwanted $\beta$-$\alpha$ and $\beta$-$^6$He pile-up events. This technique has successfully allowed to determine the decay mechanism of states in $^9$Be~\cite{yolanda03,yolanda05}, $^9$B \cite{uffe} and $^{12}$C \cite{hans}. We apply here this technique to the study of the decay mechanism of the $^{11}$Li $\beta$-delayed $^6$He$\alpha$n and 2$\alpha$3n channels, trying to determine which levels in $^{11}$Be and $^{10}$Be are involved. Moreover we discuss the possible role played by the  $^7$He and $^5$He resonances in the break-up of $\beta$-fed $^{11}$Be states.

\begin{figure}
\centerline{\epsfig{figure=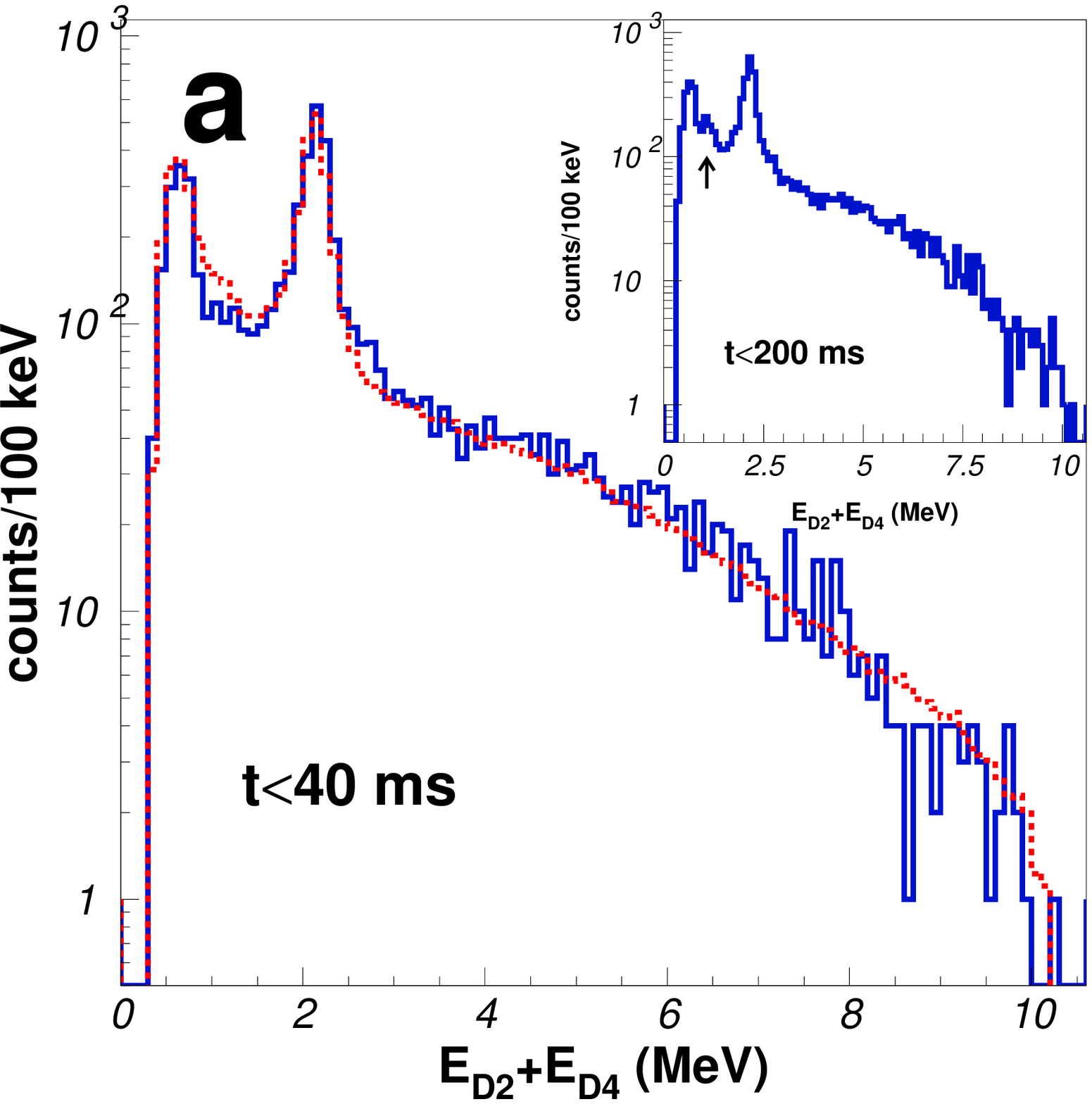,width=8cm}\epsfig{figure=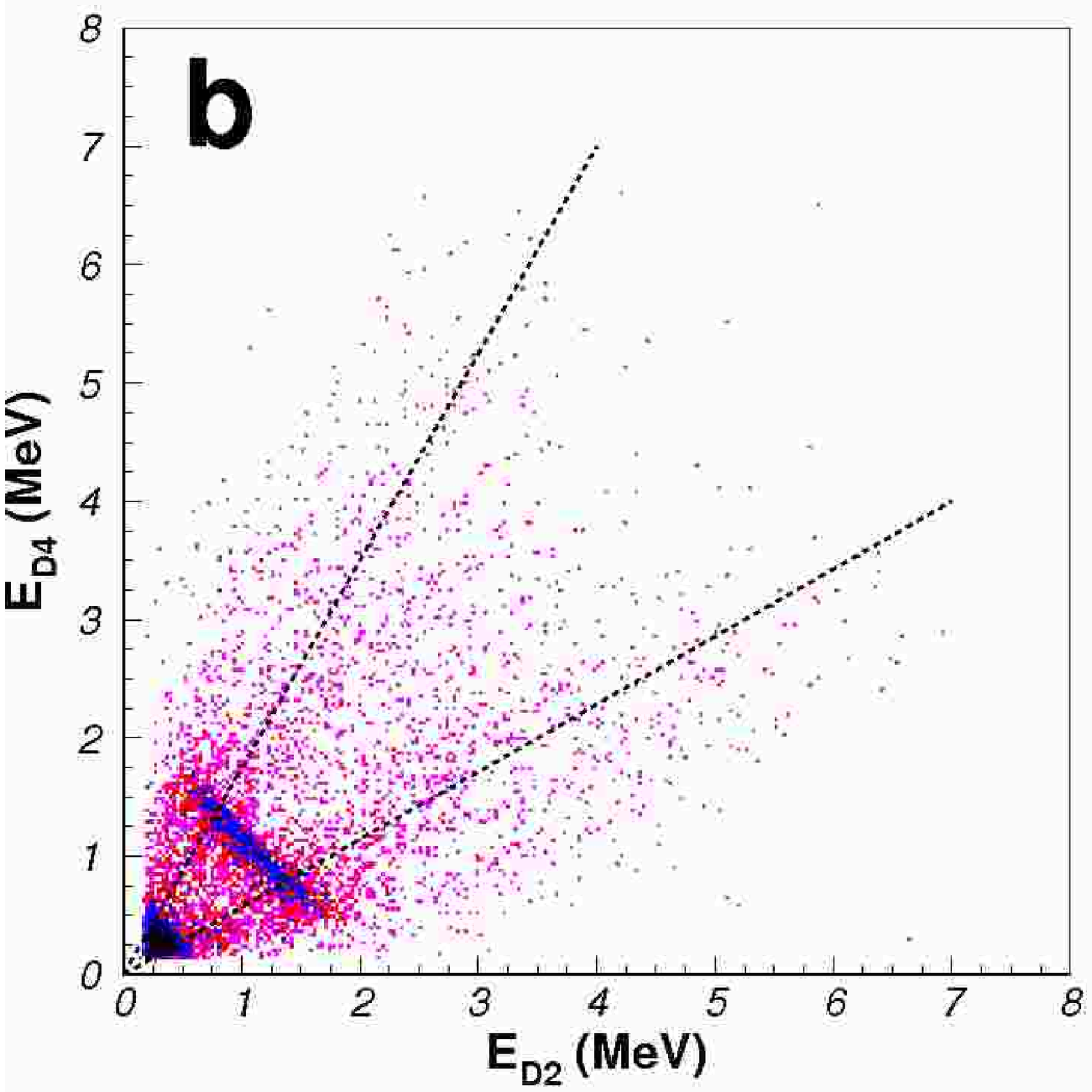,width=8cm}}
\caption{Left: The energy sum of $^{11}$Li $\beta$-delayed  coincidence spectrum collected during the first 40 ms after the proton impact. The red dashed histogram is the result of a Monte-Carlo simulation of the charged particle decay channels as described in the literature \cite{langevin,borge}. The inset shows the same sum energy spectrum for a longer time window (t$<$200 ms). The peak appearing at 1.2 MeV, indicated by an arrow, corresponds to the break-up into $^7$Li+$\alpha$  of the 9.88 MeV state in the granddaughter $^{11}$B. Right: Fig. $b$ shows the energy-vs-energy scatter plot for charged particle coincidences between the D2 and D4 opposite detectors for t$<$40 ms. The major features of the plot are the intensity at low energy and the transverse line at low energies (corresponding to peaks at 0.7 and 2.2 MeV in figure $a$). The scatter events at high energy form two lobes (at around 30 and 60 degrees).}
 \label{fig3}
\end{figure}

We concentrate here on the analysis of $^{11}$Li $\beta$-delayed double coincidences detected in the DSSSD's. The $^{11}$Li $\beta$-delayed events are enhanced over daughter activities by gating on the first 40 ms after the proton impact on the target. The combined effect of the release from the target and the half-life of $^{11}$Li (T$_{1/2}$=8.5(2) ms \cite{selovec}) makes this time window optimum for the study of the $^{11}$Li activity with minimum contribution from the $\beta$-delayed $^7$Li+$\alpha$ branch of the daughter nucleus $^{11}$Be (see Fig. \ref{fig3}$a$). The particle emitting channels following $^{11}$Li $\beta$-decay are  5-body 2$\alpha$3n, 3-body n$\alpha^6$He, 2-body $^8$Li+t and $^9$Li+d. All  these channels involve the emission of two charged fragments, but the low recoil energy of the $^{9}$Li ions ($\lesssim$ 160 keV \cite{borge}) makes the contribution of the $^9$Li+d channel negligible. The $^8$Li +t channel is sorted out from the data by kinematics. The deduced branching ratio of 0.014(3)\% is in agreement with the literature value \cite{mukha,borge}. The total number of events assigned to this decay channel, 224,  is less than 3\% of the total coincidence events. Thus, the study of the charged particle coincidences greatly enhances the 5-body and 3-body channels.

The study of the charged particle emission in $^{11}$Li  is connected with several experimental difficulties. First, we lack complete kinematics information in the five body channel, as we cannot reconstruct the three missing neutron momenta. Second, although it is possible to calculate the missing neutron momentum in the three body channel, the lack of particle identification in our setup, makes the reconstruction rather inaccurate.  Because of these shortcomings, the analysis concentrates on studying direct observables, the individual energy and direction of the detected particles and simple transformations of these energies, such as sums and differences. The sum energy spectrum enhances the contribution of the channels where the two detected particles formed  a resonance. The energy difference, defined as $E'=\frac{1}{\sqrt{2}}(E_{D2}-E_{D4})$, is the best parameter to enhance  channels where the decay kinematics group the events across the scatter plot instead of scattering them evenly, as expected from pure phase space kinematics.

\begin{table}
\begin{center}
\caption{Level centroid and reduced widths  used in the R-Matrix description of the states modeled in the Monte-Carlo code. The $\Gamma$ was obtained from a gaussian  fit of the R-matrix peak directly. \label{table1}  }
\begin{tabular}{ccccc}
 & E$_0$ (MeV) & $\gamma^2$ (MeV) & $\Gamma$ (MeV) & Ref. \\
  \hline \hline

$^{11}$Be(10.59 MeV)  &  10.59 & 0.21 & 0.227 & \cite{selover} \\
$^{11}$Be(18.15 MeV)  & 18.15 & $^{0.07~(3-body)}_{0.06~(5-body)}$ & 0.8 & \cite{borge} \\
\hline
$^{10}$Be(9.5 MeV)  &  9.52(2) & 0.21 & 0.30(4) &  This work \\
\hline
$^{7}$He(gs)  &  0.43$^{\dagger}$ & 0.4 & 0.148(1) & \cite{tilley} \\
\hline
$^{6}$He(2$^+$)  & 1.8 & 0.113  & 0.117(1) & \cite{tilley} \\
\hline
$^{5}$He(gs)  & 0.895$^{\ddagger}$ & 2.5 &  0.658(4) & \cite{tilley} \\
\hline\hline
 \end{tabular}
\end{center}
$^{\dagger}$ above the $^6$He+$\alpha$ threshold. \\
$^{\ddagger}$ above the $\alpha$+n threshold.

\end{table}

%{\bf The sum energy spectrum enhances the contribution of the channels where the two detected particles belongs to a resonance. On the other hand, the energy difference spectrum, defined for energies $E'=\frac{1}{\sqrt{2}}(E_{D2}-E_{D4})$, is the best method to distinguish between phase space or sequential emission.}{\it Miguel and Maria: Here we are not sure what you meant and you have to check if you agree that this is the method to unravel the decay mechanism?}

Fig. \ref{fig3}$a$ shows the sum energy spectrum corresponding to the coincidences observed in the two opposite detectors D2 and D4 for a time interval t$<$40 ms. The sum energy spectrum compares well with the one previously measured (see Fig. 10 of Ref. \cite{langevin}). The 0.7 and 2.2 MeV peaks were assigned \cite{langevin} to the break-up of the 10.59 MeV state in $^{11}$Be \cite{selover} into 5-body (2$\alpha$3n) and 3-body (n$\alpha^6$He) channels, through an intermediate state in $^{10}$Be around 9.5 MeV, see Fig. \ref{fig1}. The continuous distribution at higher energy was assigned to the  5-body break-up of the 18.15(15) MeV state in $^{11}$Be \cite{langevin,borge}.  In the inset of Fig. \ref{fig3}$a$ the sum energy spectrum for the first 200 ms after the proton impact is shown. For this longer time interval a low intensity peak at around 1.2 MeV appears,  corresponding to the decay channel of the daughter nucleus $^{11}$Be~$\stackrel{\beta}{\longrightarrow}$~$^{11}$B$^*\longrightarrow$$^7$Li+$\alpha$ (T$_{1/2}$=13.81(8) s \cite{selovec}). This channel was used for normalization of the branching ratios as explained later on.

\begin{figure}
\centerline{\epsfig{figure=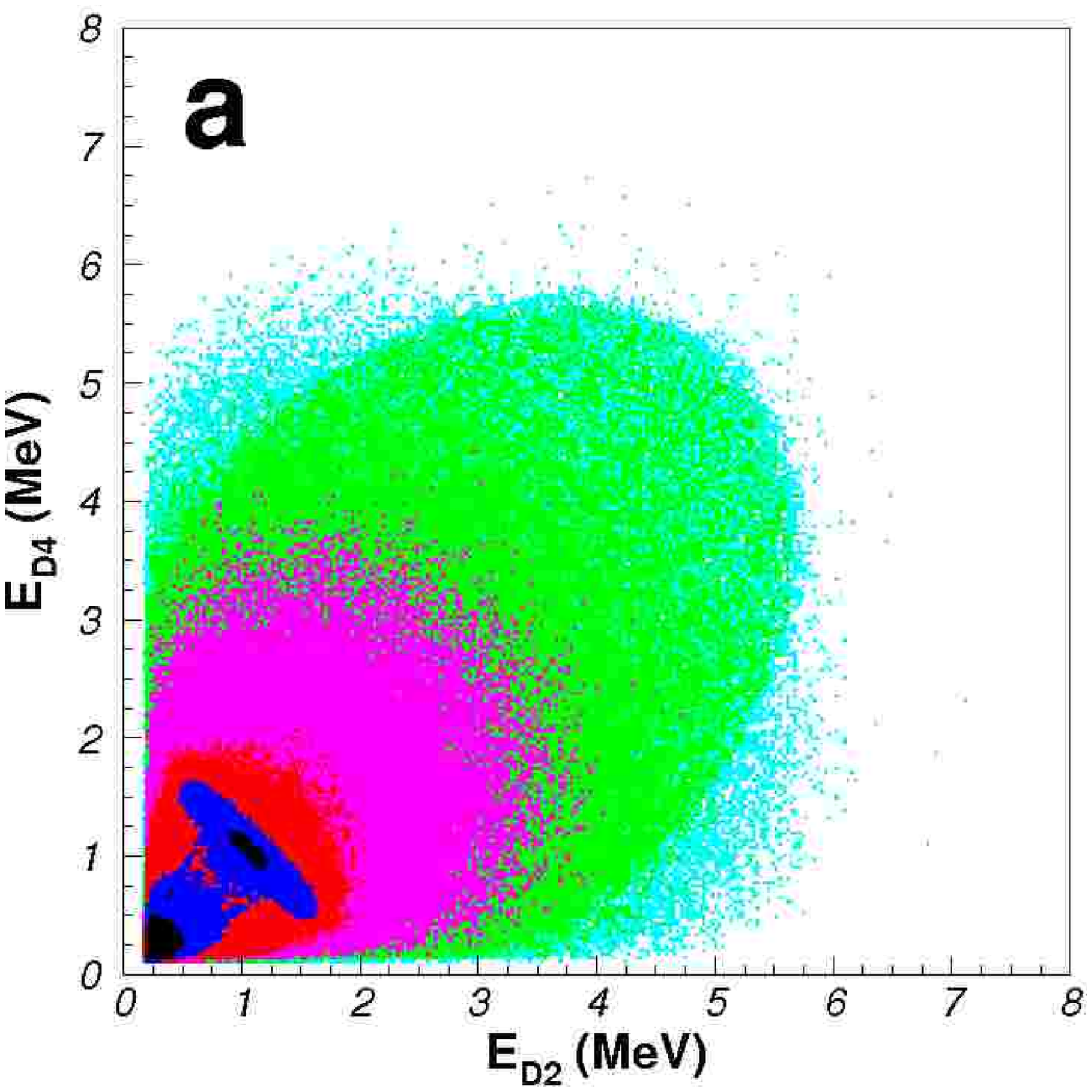,width=8cm}\epsfig{figure=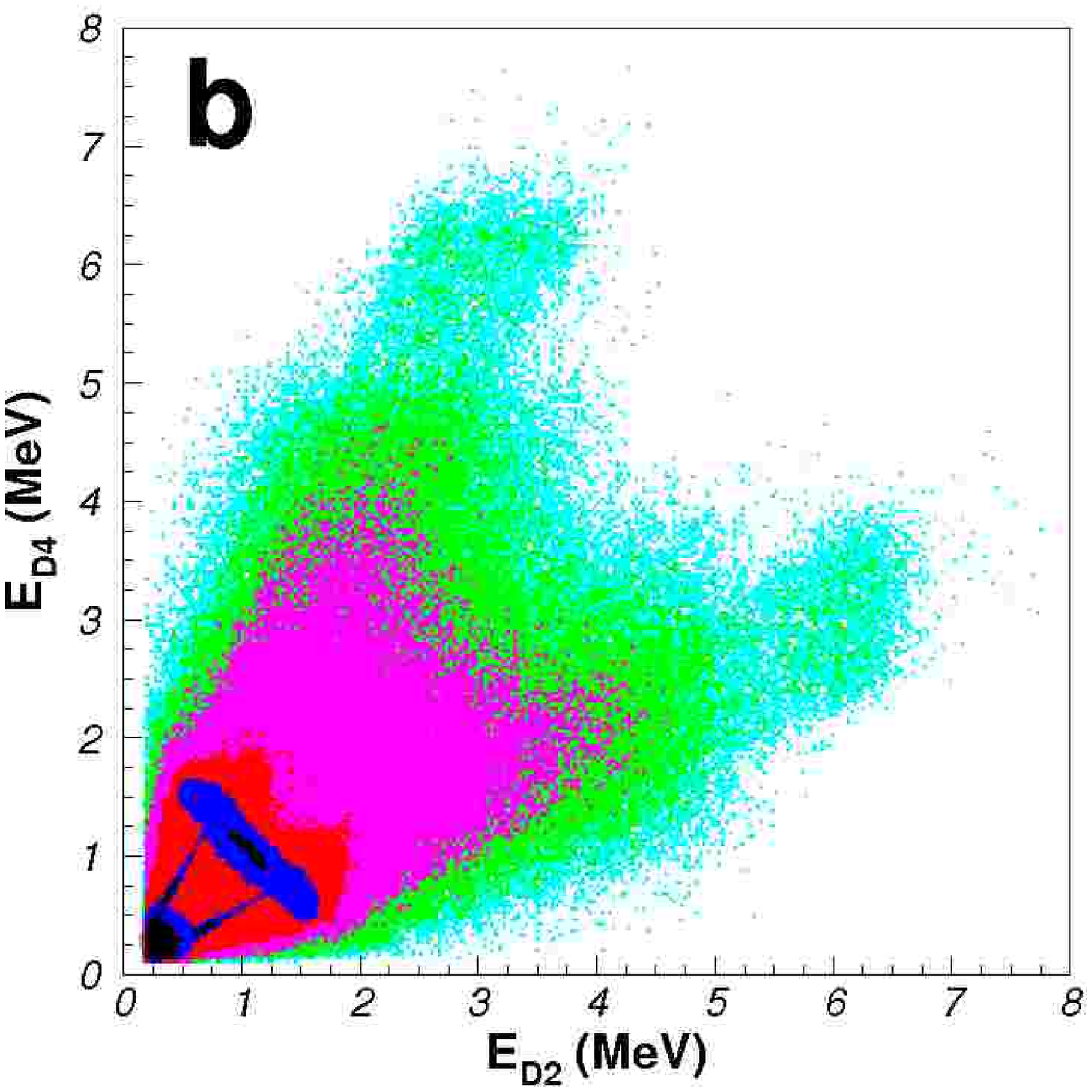,width=8cm}}
\caption{Left: Monte-Carlo simulation of the expected E$_{D2}$ vs E$_{D4}$ scatter plot of the $^{11}$Li $\beta$-delayed charged particles assuming the decay channels proposed in Refs. \cite{langevin,borge}, and schematically shown in Fig. \ref{fig1}. The main features are reproduced. However, the high energy distribution, modeled as phase-space of the $^{11}$Be(18.15 MeV) level breaking into 3-body and 5-body particles, is very different from the experimental one shown in Fig. \ref{fig3}$b$. In particular there are very few events with charged particles of high energies. Right: Fig. $b$ shows the E$_{D2}$ vs E$_{D4}$ scatter plot of the Monte-Carlo simulation of the $^{11}$Li $\beta$-delayed charged particle break-up including the new channels through He isotopes listed in Table \ref{table2}.}
\label{fig4}
\end{figure}

The red dashed histogram in Fig \ref{fig3}$a$ shows a Monte-Carlo simulation of the charged particle energy distribution assuming the previously described decay channels (Fig. \ref{fig1}). The detector efficiency for each channel is included in the simulation. The $^{10}$Be and $^{11}$Be levels are described using R-Matrix theory following the approximation given in the Appendix of Ref. \cite{nyman}. The parameters of the 10.59 MeV state in $^{11}$Be were taken from the $^9$Be(t,p) reaction \cite{selover} while those for the 18.15 MeV state were taken from \cite{borge}. The study of the three body n$\alpha^6$He  break-up channel through $^{10}$Be$^*$ states allows a determination of the energy and width of this intermediate state from a fit of the relative energy of the detected $^6$He and $\alpha$ particle. The energy and width of the intermediate state in $^{10}$Be obtained from the fit are 9.52(2) and 0.30(4) MeV respectively. Previous reaction studies indicated the existence of excited states in $^{10}$Be in this energy region. In neutron exchange reaction an state is identified  with energy and width values of 9.4 MeV and 290(20) keV respectively \cite{anderson}. In a recent Li($^7$Li,$\alpha^6$He)$\alpha$ reaction study \cite{curtis} a level is found at 9.56(2) MeV excitation energy and 141(10) keV width. Including a narrow state with the latter values in our simulation deteriorates the fit considerably. Due to the difference between our results and the one of  reference \cite{curtis} we cannot conclude that it is the same level. The different parameters used in the Monte-Carlo simulation are given in Table \ref{table1}.

\begin{table}
\caption{$\beta$-fed states in $^{11}$Be above the charged particle thresholds and their break-up channels. \label{table2}}
\begin {tabular}{p{1.7cm}p{2.1cm}p{1.9cm}|p{1.8cm}p{2.4cm}p{1.5cm}}
\multicolumn{3}{l}{\footnotesize{Decay channels proposed previously.}} & \multicolumn{3}{l}{\footnotesize{Decay channels from this work.}}  \\
\hline\hline
  \multirow{2}{*}{$^{11}$Be(10.59)}  & \multirow{2}{*}{$\stackrel{n}{\longrightarrow}$$^{10}$Be(9.5)}  & $\stackrel{\alpha}{\longrightarrow}$\footnotesize{$^6$He} \footnotesize{\cite{langevin}} &   \multirow{3}{*}{$^{11}$Be(10.59)}  & \multirow{2}{*}{$\stackrel{n}{\longrightarrow}$$^{10}$Be(9.5)} &  $\stackrel{\alpha}{\longrightarrow}$$^6$He \\
 & &   $\rightarrow$\footnotesize{ 2$\alpha$3n} \footnotesize{\cite{langevin}} &  &  & $\rightarrow$ 2$\alpha$3n  \\
 & & &   & $\stackrel{\alpha}{\longrightarrow}$$^{7}$He(gs) &  $\stackrel{n}{\longrightarrow}$$^6$He \\ \hline
  \multirow{2}{*}{$^{11}$Be(18.15)} & $\longrightarrow$ 2$\alpha$3n \cite{langevin} &   &\multirow{2}{*}{ $^{11}$Be(18.15)}  & \begin{scriptsize}$\longrightarrow$$^6$He(2$^+$)+$^5$He(gs)\end{scriptsize} &  $\rightarrow$ 2$\alpha$3n$^{\dagger}$ \\
 & \footnotesize{$\rightarrow$ n$\alpha$$^6$He} \cite{borge} &   & & $\stackrel{\alpha}{\longrightarrow}$$^{7}$He(gs) &  $\stackrel{n}{\longrightarrow}$$^6$He \\
\hline\hline
\end {tabular}
$^{\dagger}$ Up to a 20\% admixture of the $^{11}$Be(18.15)$\rightarrow$2$\alpha$3n direct channel  
cannot be excluded (see text).

\end{table}

Fig. \ref{fig3}$b$ shows the scatter plot corresponding to coincidence events observed in the opposite detectors D2 and D4 for times t$<$40 ms. The most significant features of the plot are the intensity at low energy and the accumulation of events forming a transverse line at low energy, corresponding to the two peaks at 0.7 and 2.2 MeV in the sum energy spectrum. Careful inspection of the high energy region reveals that the coincidences are not evenly distributed, they rather form two elongated bumps or lobes, highlighted by the black dotted lines. Comparison of the experimental scatter plot of Figs. \ref{fig3}$b$ and the Monte-Carlo simulation, shown in Fig. \ref{fig4}$a$ indicates that although the main features of the data are reproduced by the simulation, the distribution of the high energy coincidences is not. This is not a surprise, as the 5-body and 3-body break-up of the 18.15 MeV state in $^{11}$Be are simulated using a phase space momentum distribution, which neglects the role of possible structure effects.

%9.5 width is 0.276(40)

As just noted, the incline of the lobes in the scatter plot of Fig. \ref{fig3}$b$ corresponds to lines of slope 7/4 and 4/7. This structure corresponds to a mass-asymmetric two-body break-up where the initial state is broad. The only mass 4 and mass 7 isobars energetically allowed for $^{11}$Li $\beta$-decay are $^7$He+$\alpha$. This comes naturally, as alpha emission is common to all Be isotopes. Furthermore, the $^7$He(gs) is the only resonance in $^7$He below the $\alpha$+3n threshold, see Fig. \ref{fig1}, thus it is a natural choice as intermediate state in the 3-body break-up of $^{11}$Be, i.e. $^{11}$Be(18.15 MeV)$\rightarrow\alpha+^7$He(gs)$\rightarrow$n+$^6$He+$\alpha$. The energy distribution of this decay channel appears as two body break-up, it keeps memory of the first step if this Q value is much larger than the second one.  It is known that the $^7$He(gs) is situated 430 keV above the $^6$He+n threshold \cite{tilley}, therefore the recoil energy given by the emission of the neutron will be low (61 keV). Thus, the distribution of the two detected charged particles, $\alpha$ and $^6$He, in the scatter plot will kinematically be very close to a two body break-up, but broadened by the recoiling neutron. Other possible intermediate steps involving the ground state of $^6$He and the $^5$He(gs) resonance would also end up in the 3-body n$\alpha^6$He channel, but in this case the slope of the two bands would be 6/5 and 5/6, quite different from the experimental distribution.

\begin{figure}
\centerline{\epsfig{figure=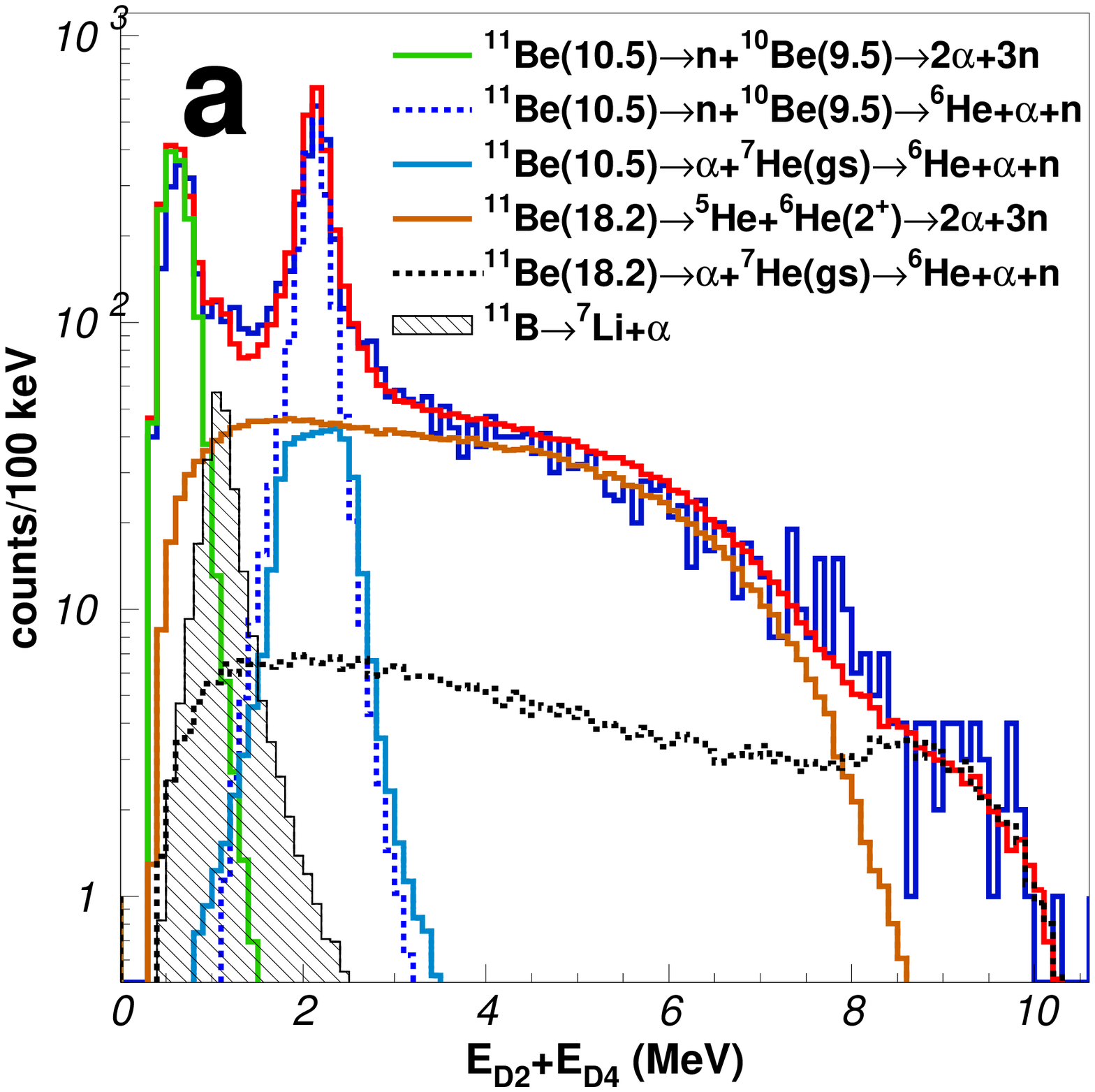,width=8cm}\epsfig{figure=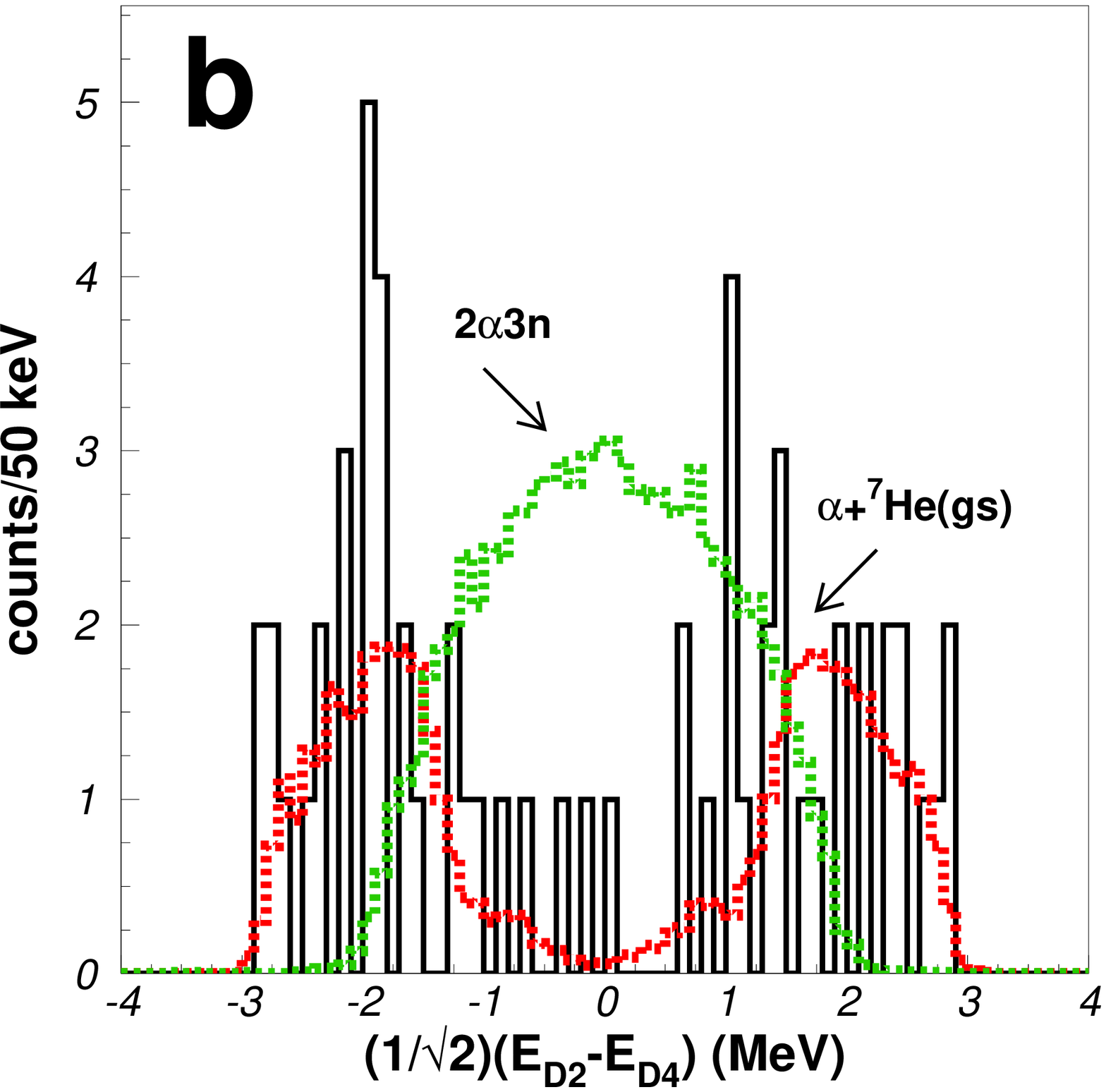,width=8cm}}
\caption{Left: Sum energy spectrum in blue,  Monte-Carlo simulation including the $^A$He channels in red. The different break-up channels are also shown, with color code according to the legend. Right:  Spectrum of the energy difference, $E'$, for events with 8.2$\leq$E$_{sum}$$\leq$10 MeV compared to two Monte-Carlo simulations:  in red the one including the $^7$He(gs) and in green the one corresponding to the 5-body phase-space break-up channel. The contribution of the $^7$He(gs) channel is necessary to successfully reproduce the central dip in the experimental distribution. }
\label{fig5}
\end{figure}

It is important here to point out that the 3-body break-up of the $^{11}$Be states through the ground state of $^7$He cannot be the only decay channel, as in previous experiments a factor of roughly 5 $\alpha$ particles per $^6$He ion was determined \cite{langevin2}. This implies that the majority of the high energy events correspond to the 2$\alpha$3n channel. However, the phase space description of the 2$\alpha$3n channel is not able to explain the energy distribution of coincidence events. On the contrary, if sequential decay is assumed, the first step could be  $^6$He(2$^+$)+$^5$He(gs), as both states are above the $\alpha$+2n and $\alpha$+n thresholds, resulting in the 2$\alpha$3n final channel. This channel is a very good candidate, as the states are broad, widths of 0.113 and 0.648 MeV respectively \cite{tilley}, spreading the coincidence data in a similar way as the 5-body phase space distribution, but with a dip in the central part, E$_{D2}\simeq$~E$_{D4}$, as observed in the scatter plot of Fig. \ref{fig3}$b$.

A new Monte-Carlo simulation was performed. For the 10.59 MeV state in $^{11}$Be the two break-up channels from the literature \cite{langevin} were included. Furthermore, the  break-up of the 10.59 MeV state into $\alpha$+$^7$He(gs) was considered as a possible contributing channel, since the fingerprint of this channel was seen near the transverse line (see Fig. \ref{fig3}$b$). Although this channel is expected to contribute with ten times less intensity  than the decay through $^{10}$Be, the $\chi^2$ fit of the sum energy spectrum improves by 20\% if this channel is included. For the break-up of the 18.15 MeV state in $^{11}$Be the new channels involving the $^{5-7}$He resonances were considered. The $\chi^2$ analysis of the energy difference spectra indicates that up to a 20\% admixture of the 5-body  direct break-up channel is statistically equivalent to the simulation without any phase space contribution. Table \ref{table1} lists the parameters used to describe the various resonances involved and Table \ref{table2} summarizes the different decay channels included in this simulation. The resulting sum energy spectrum is shown Fig. \ref{fig5}$a$. The experimental data and the Monte-Carlo simulation of the proposed channels are in excellent agreement. There is a minor discrepancy between the total simulation and the experimental spectrum at around 7.8 MeV. This could suggest the contribution of another state in $^{11}$Be, as previously proposed \cite{borge,fynbo04}, but the lack of statistics in this region prevents us from finding a conclusive explanation for this minor discrepancy. The great improvement achieved by including the sequential decay channels through the $^A$He resonances is shown in the comparison of the new  scatter plot shown in Fig. \ref{fig4}$b$ with the experimental one of Fig. \ref{fig3}$b$. This is further supported by an excellent fit of the coincidence energy spectrum.

\begin{table}
\begin{center}
\caption{Branching ratios for channels determined in this work following $^{11}$Li $\beta$-decay. The total branching ratio to charged particle emitting channels obtained in \cite{langevin} is 3.1(9)\%, compared to the value of 1.73(2)\%. in our work. The $^{11}$Li activity was deduced from the branching of the $\beta$-($^7$Li+$\alpha$) decay channel of the daughter. \label{table3} }
\begin{tabular}{lccc}
 Channel & $\beta$-feeding (\%)$^{\dagger}$  & $\beta$-feeding (\%)$^{\ddagger}$ & $\beta$-feeding (\%)$^{\S}$  \\
\hline\hline
\scriptsize{$^{11}$Be(10.59)$\rightarrow$n+$^{10}$Be(9.5)$\rightarrow$2$\alpha$+3n} &  1.08(2)  & 1.1(2)  & 1.4(2) \\
\scriptsize{$^{11}$Be(10.59)$\rightarrow$n+$^{10}$Be(9.5)$\rightarrow$n+$\alpha+^6$He} & 0.227(5)  & 0.23(4)  & 0.29(4)  \\
\scriptsize{$^{11}$Be(10.59)$\rightarrow\alpha+^7$He$\rightarrow$n+$\alpha+^6$He}  & 0.0348(5)  &    0.035(6)  & 0.044(7)      \\
\scriptsize{$^{11}$Be(18.15)$\rightarrow$$^6$He(2$^+$)+$^5$He$\rightarrow$2$\alpha$+3n} & 0.337(7)  & 0.34(5)  & 0.43(7)   \\
\scriptsize{$^{11}$Be(18.15)$\rightarrow\alpha+^7$He$\rightarrow$n+$\alpha+^6$He}  & 0.057(1) &    0.057(9)  & 0.072(10)     \\
\hline\hline
\end{tabular}
\end{center}
$^{\dagger}$ Only the statistical error is consider in this column. \\
$^{\ddagger}$ Including the normalization uncertainty. \\
$^{\S}$ Assuming a 2\% feeding to the ground state in $^{11}$Be, stated as upper limit in previous works \cite{bjornstad}. \\

\end{table}

The  energy difference distribution, $E'$, in the 8.2$<$E$_{D2}$+E$_{D4}$$<$10 MeV region, is shown in the black histogram in Fig. \ref{fig5}$b$. As mentioned before, the $E'$ parameter enhances the decay channels where the detected particles are one from the first step and the other from the final resonance. We chose this high energy region because one or maximum two channels could contribute to the spectrum. The phase-space break-up Monte-Carlo simulation is shown in dashed green in Fig. \ref{fig5}$b$, while the sequential break-up including the $^7$He(gs) channel is shown in dashed red. The data show a distribution with a clear dip in the middle that is contradicted by the phase-space model. On the other hand, the inclusion of the $^7$He+$\alpha$ channel nicely reproduces the double peak shape of the plot. This region is clearly dominated by the $^{7}$He(gs) channel  (see  the black dashed line in Fig. \ref{fig5}$a$).

Table \ref{table3} shows the branching ratios to levels in $^{11}$Be obtained in this work. The values were calculated from the intensity distribution of the Monte-Carlo simulation, rather than from the direct experimental data, as the experimental coincidence efficiency is difficult to determine. In order to normalize the $\beta$-branching ratios we need to estimate the absolute $^{11}$Li activity. The  activity was evaluated using the known branching ratio of the daughter, $^{11}$Be, decay into $^7$Li(gs and 1/2$^-$)+$\alpha$. The $^{11}$Li $\beta$-decay branching ratio to the first excited state in $^{11}$Be, 7.4(3) \%, is the weighted average of the values from \cite{mjborge,bjornstad,aoi,morrissey,detraz}. The value for $^{11}$Be $\beta$-decay branching ratio to the 9.8 MeV state in $^{11}$B is 2.9(4) \% \cite{alburger}. The values shown in the second column in Table \ref{table3} were obtained assuming negligible $\beta$-feeding to the $^{11}$Be ground state. The third column shows the branching ratios obtained assuming 2\% $\beta$-feeding to the $^{11}$Be ground state. A possible source of uncertainty in the estimation of the $\beta$-feeding values  is the presence of $^{11}$Be ions in the incoming $^{11}$Li beam. Direct production of $^{11}$Be up to a few percent of the $^{11}$Li beam has been observed previously at ISOLDE \cite{mjborge}. If this is the case, the branching ratios will increase in the same few percent as the direct production of $^{11}$Be.

Comparison with previous branching ratio values \cite{langevin} is not considered pertinent. Although the energy and width of the states in $^{11}$Be are taken from the literature \cite{selover,borge}, the break-up channels are different in this and previous analysis, thus resulting in different distribution of the $\beta$ feeding. As an additional check, we compare the  branching ratio of the 2$\alpha$+3n channel to the published  P$_{3n}$ value, 1.9(2)\% \cite{azuma,mjborge}, obtained assuming no $\beta$-feeding to the ground state. The value obtained in this work, 1.4(5)\%, is smaller, but one has to take into account  that the 2$\alpha$3n breakup of the 10.59 MeV state has a distribution that is below our particle threshold.

\section{Summary and conclusions}

The charged particles emitted in the $\beta$-decay of $^{11}$Li have been measured in coincidence using a highly segmented setup. The coincidence energy spectrum can be explained mainly by the 2$\alpha$+3n and n+$\alpha$+$^6$He channels. The analysis presented in this work shows that the  phase space energy description of the 2$\alpha$3n channel is not capable of explaining the correlations at high energy observed in the $\beta$-decay of $^{11}$Li (Fig. \ref{fig3}$b$). In particular the lack of events in the central region of the scatter plot is intriguing. This was also visible in Fig. 9 of Ref. \cite{langevin} although not considered by the authors, probably due to the low statistics. This minimum forms two lobes of slopes 7/4 and 4/7 indicating the involvement of mass 4 and mass 7 nuclei in the first stage of a sequential decay. This naturally brings  $^A$He isotopes into play, as they are involved in the break-up of all other Be isotopes. A Monte-Carlo simulation was performed including these channels, $^{11}$Be(10.59, 18.15)$\rightarrow~ \alpha+^7$He~$\rightarrow~^6$He$+\alpha+$n and $^{11}$Be(18.15)$\rightarrow ^6$He$(2^+)+^5$He$\rightarrow2\alpha+3$n, plus the confirmed n$\alpha^6$He break-up of the 10.59 MeV state in $^{11}$Be through the 9.5 MeV state in $^{10}$Be, whose energy and width have been determined in this work. The resulting simulation reproduces well the correlations observed in the scatter plot,  thus rejecting a description of the break-up of the 18.15 MeV state in terms of phase space energy sharing only, although up to 20\% mixture of direct break-up can not be statistically excluded. We also check the energy difference spectrum E$_{D2}$-E$_{D4}$ in the region where we expect the $^7$He+$\alpha$ sequential decay channel to dominate. Neither 5-body 2$\alpha$+3n break-up \cite{langevin} nor  3-body n$\alpha^6$He break-up \cite{borge}  described by  phase space energy distribution reproduces the structure observed in this spectrum. Only by introducing the sequential  break-up through  $^7$He(gs) into n$\alpha^6$He final state we are able to explain the two-peak structure observed in the energy difference distribution, $E'$.

The study of the $^{11}$Li $\beta$-delayed particle emission is an excellent tool to determine the highest excited states in $^{11}$Be,  a difficult task to be singled out by other experimental techniques. These results constitute a significant step beyond the previous $\beta$-decay work \cite{langevin}. The experimental data presented in this work indicate that the decay of excited states in $^{11}$Be above the charged particle thresholds cannot be explained solely as a multi-particle break-up dominated by the phase space energy distribution. We conclude that $^A$He resonant states play a significant role in the sequential break-up of $^{11}$Be excited states. This naturally suggests a more prominent role of nuclear structure effects in the description of the break-up process of these states than previously assumed.

\section{Acknowledgements}

This work has been supported by the  Spanish CICYT, under the projects FPA2002-04181-C04-02, FPA2005-02379 and the MEC Consolider project \linebreak CSD2007-00042, the European Union Sixth Framework through RII3-EURONS (contract no. 506065) and the Swedish Knut and Alice Wallenberg Foundation. M. Madurga acknowledges the support of the Spanish MEC under the FPU program, FPU AP-2004-0002. T. Nilsson is a Royal Swedish Academy of Sciences Research Fellow supported by a grant from the Knut and Alice Wallenberg Foundation. We acknowledge the help and support of the ISOLDE Collaboration.

\clearpage

\clearpage

\clearpage

\clearpage

\clearpage

\clearpage

\clearpage

\clearpage

\end{document}